\documentclass [a4paper,fleqn, 12pt]{article}
\usepackage{graphicx}
\usepackage[small]{subfigure,epsfig}

\usepackage {amsmath} \usepackage{amssymb} \usepackage{cite}

\newcommand{\pr}{\textbf{Proof.}\ }
\newcommand{\cn}

\begin{document}


\title
{On elliptic solutions of nonlinear ordinary differential equations}

\author
{Maria V. Demina, \and Nikolay A. Kudryashov}

\date{Department of Applied Mathematics, National Research Nuclear University
MEPHI, 31 Kashirskoe Shosse,
115409 Moscow, Russian Federation}




\maketitle

\begin{abstract}

The generalized Bretherton equation is studied. The classification of the meromorphic traveling wave solutions for this equation is presented. All possible exact solutions of the generalized Brethenton equation are given.

\end{abstract}






\section{Introduction}

At present there exists a lot of methods for finding exact elliptic
solutions of autonomous nonlinear ordinary differential equations.
Let us name only a few: the Weierstrass function method \cite{Kudr90a, Kudr91}, the Jacobi elliptic--function method \cite{Parkes02, Fu02, Kudr05} and their
different extensions and modifications \cite{Fan00, Kudr92, Parkes01, Biswas01, Bagderina01, Kudr06, Kudr08}. Making use of such a method one can not but come
across the following questions. Whether all families of elliptic
solutions are found. Whether an equation does not have elliptic
solutions at all, if a method fails to find any. This questions are
addressed very seldom (nevertheless see \cite{Vernov02, Demina01, Demina 02}).

The aim of this article is to present an algorithm, which enables
one to find all families of doubly periodic meromorphic solutions
satisfying an autonomous nonlinear ordinary differential equation.
We will use an approach suggested in \cite{Demina01, Demina02}. As
an example we take the following third order differential equation
\begin{equation}
\begin{gathered}
\label{Example_introduction}w_{zzz}+3\eta w w_{zz}+\eta w_z^2+\beta
w_{zz}+2\alpha w^3+\delta ww_z+\gamma w_z +\sigma w^2+\mu w+ \nu =0
\end{gathered}
\end{equation}
with $\eta\neq 0$, $\alpha \neq 0$. We find all conditions for
elliptic solutions to exist and construct elliptic solutions in
explicit form.

This article is organized as follows. In section $2$ we present our
method and give explicit expressions for elliptic solutions of an
autonomous nonlinear ordinary differential equation. In section  $3$
we consider an example and classify elliptic solutions of equation
\eqref{Example_introduction}.

\section{Method applied} \label{Method applied}

Consider an algebraic autonomous nonlinear ordinary differential
equation
\begin{equation}
\label{EQN} E[w(z)]=0.
\end{equation}
Let us look for its elliptic solutions. If $w(z)$ is such a
solution, then equation \eqref{EQN} has the family of elliptic
solutions $w(z-z_0)$ with $z_0$ being an arbitrary constant.
Equation \eqref{EQN} necessary possesses an elliptic solution, if it
admits at least one Laurent expansion in a neighborhood of the pole
$z=z_0$.  Without loss of generality we may build Laurent series  in
a neighborhood of the point $z=0$
\begin{equation}
\begin{gathered}
\label{LE1}
w(z)=\sum_{k=1}^{p}\frac{c_{-k}}{z^k}+\sum_{k=0}^{\infty}c_kz^k,\quad
0<|z|<\varepsilon.
\end{gathered}
\end{equation}
Here $p$ is the order of the pole $z=0$.

\textbf{\textit{Proposition.}} \textit{ Suppose Laurent series
\eqref{LE1} with uniquely determined coefficients  satisfies
equation \eqref{EQN}, then this equation admits at most one
meromorphic solution having a pole $z=0$ with Laurent series
\eqref{LE1}.}

This proposition follows from the propertied of Laurent series and
uniqueness of analytic continuation. As a consequence, equation
\eqref{EQN} may have at most one elliptic solution possessing the
pole $z=0$ with Laurent series \eqref{LE1}. The order of an elliptic
function is defined as the number of poles in a parallelogram of
periods, counting multiplicity.

Our algorithm for finding elliptic solutions of equation \eqref{EQN}
in closed form is the following. Note that we omit arbitrary
constant $z_0$.

\textit{Step 1.} Perform local singularity analysis around movable
singular points for solutions of equation \eqref{EQN}.

\textit{Step 2.} Select the order $M$ of $w(z)$ and take $K$
distinct Laurent series
\begin{equation}
\begin{gathered}
\label{LS_K}
w^{(i)}(z)=\sum_{k=1}^{p_i}\frac{c_{-k}^{(i)}}{z^k}+\sum_{k=0}^{\infty}c_k^{(i)}z^k,\quad
0<|z|<\varepsilon_i,\quad i=1, \ldots, K.
\end{gathered}
\end{equation}
from those, found at step 1, in such a way that the following
conditions
\begin{equation}
\begin{gathered}
\label{ElS_Residues} \sum_{i=1}^{K}c_{-1}^{(i)}=0,\quad
\sum_{i=1}^{K}p_i=M.
\end{gathered}
\end{equation}
hold.

\textit{Step 3.} Define the general expression for the elliptic
solution $w(z)$ possessing $K$ poles $a_1$, $\ldots$,  $a_K$ in a
parallelogram of periods (see \cite{Lavrentiev01, Demina01, Demina
02}). The Laurent expansion in a neighborhood of the point $z=a_i$
is $w^{(i)}(z-a_i)$. In other words, take the following expression
for $w(z)$
\begin{equation}
\begin{gathered}
\label{Ex_Sol_EllipticK}
w(z)=\sum_{i=1}^{K}c_{-1}^{(i)}\zeta(z-a_i)+\left\{\sum_{i=1}^{K}
\sum_{k=2}^{p_i}\frac{(-1)^k
c_{-k}^{(i)}}{(k-1)!}\frac{d^{k-2}}{dz^{k-2}}\right\}\wp(z-a_i)
+\tilde{h}_0.
\end{gathered}
\end{equation}
Here $\wp(z)$ is the Weierstrass function satisfying the equation
\begin{equation}
\begin{gathered}
\label{Wier} (\wp_z)^2=4\wp^3-g_2\wp-g_3,
\end{gathered}
\end{equation}
$\zeta(z)$ is the Weierstrass $\zeta$--function, $\tilde{h}_0$ is a
constant.

\textit{Step 4.} Find the Laurent series for $w(z)$ given by
\eqref{Ex_Sol_EllipticK} around its poles $a_1$, $\ldots$, $a_K$.
Without loss of generality set $a_1=0$. Introduces notation
$A_i\stackrel{def}{=}\wp(a_i)$, $B_i\stackrel{def}{=}\wp_z(a_i)$,
$i=2$, $\ldots$, $K$, and
\begin{equation}
\begin{gathered}
\label{h0} h_0\stackrel{def}{=}\tilde{h}_0-
\sum_{i=2}^{K}c_{-1}^{(i)} \zeta(a_i)-\sum_{i\in \,I}^{}c_{-2}^{(i)}
\wp(a_i).
\end{gathered}
\end{equation}
With the help of addition formulae for the functions $\wp$ and
$\zeta$ (see \cite{Lavrentiev01, Demina01, Demina 02}) rewrite
expression \eqref{Ex_Sol_EllipticK} as
\begin{equation}
\begin{gathered}
\label{Ex_Sol_EllipticK_AB}
w(z)=\left\{\sum_{i=2}^{K}\sum_{k=2}^{p_i}\frac{(-1)^k
c_{-k}^{(i)}}{(k-1)!}\frac{d^{k-2}}{dz^{k-2}}\right\}\left(\frac14\left[
\frac{\wp_z(z)+B_i}{\wp(z)-A_i}\right]^2-\wp(z)\right)\\
+\sum_{i=2}^{K}\frac{c_{-1}^{(i)}(\wp_z(z)+B_i)}{2\,(\wp(z)-A_i)}+\left\{\sum_{k=2}^{p_{1}}\frac{(-1)^k
c_{-k}^{(1)}}{(k-1)!}\frac{d^{k-2}}{dz^{k-2}}\right\}\wp(z)+ h_0,
\end{gathered}
\end{equation}

\textit{Step 5.} Compare coefficients of the series found at the
second and the fourth steps. Form a system of algebraic equations.
Add to this system equations $B_i^2=4A_i^3-g_2A_i-g_3$, $i=2$,
$\ldots$, $K$. The number of equations in the system is slightly
more than the number of parameters of elliptic solution
\eqref{Ex_Sol_EllipticK_AB} and equation \eqref{EQN}. Solve the
algebraic system for the parameters of the elliptic solution $w(z)$,
i.e. find $h_0$, $g_2$, $g_3$, $A_i$, $B_i$, $i=2$, $\ldots$, $K$.
In addition correlations for the parameters of equation \eqref{EQN}
may arise. If this system is inconsistent, then equation \eqref{EQN}
does not possess elliptic solutions with supposed Laurent expansions
around poles.

\textit{Step 6.} Check-up obtained solutions, substituting them into
original equation.

In the case $g_2^3-27g_3^2=0$ the elliptic function $\wp(z)$
degenerates and consequently elliptic solution
\eqref{Ex_Sol_EllipticK_AB} degenerates. The invariants $g_2$, $g_3$
are related with the half-period $\omega_1$, $\omega_2$ by means of
equalities
\begin{equation}
\begin{gathered}
\label{Invariants} g_2={\sum_{(n,\,m)\neq(0,\,0)}}
\frac{60}{(2n\omega_1+2m\omega_2)^4},\,
g_3={\sum_{(n,\,m)\neq(0,\,0)}}
\frac{140}{(2n\omega_1+2m\omega_2)^6}.
\end{gathered}
\end{equation}

With the help of presented algorithm one can construct any elliptic
solution of equation \eqref{EQN}. Note that if equation \eqref{EQN}
possesses only $N$ distinct Laurent series in a neighborhood of
poles, then the orders of its elliptic solutions is not more than
$\sum_{i=1}^{N}p_i$, where $p_i$ ($i=1$, $\ldots$, $N$) are the
orders of poles \cite{Demina01, Demina 02}. Thus we see that our
approach may be used, if one need to classify families of elliptic
solutions satisfying equation \eqref{EQN}.

\section{Elliptic solutions of an autonomous third order ordinary differential equation}

As an example let us classify doubly periodic meromorphic solutions
of the third order nonlinear ordinary differential equation
\eqref{Example_introduction}. Without loss of  generality we may set
$\eta =1$, $\sigma=0$ to obtain
\begin{equation}
\begin{gathered}
\label{Example_THO}w_{zzz}+3w w_{zz}+w_z^2+\beta w_{zz}+2\alpha
w^3+\delta ww_z+\gamma w_z+\mu w+ \nu =0,
\end{gathered}
\end{equation}
where $\alpha \neq 0$. Our results are gathered in theorem \ref{T3}.

\textbf{Theorem 1.}
\label{T3} Equation \eqref{Example_THO} admits exact elliptic
solutions if and only if
\newline
\begin{equation}
\begin{gathered}
\label{Example_THO_P1}\text{Case I}:\quad
\delta=\frac{12}{11}\alpha,\quad \gamma=-\frac{3}{22}\alpha\beta
\end{gathered}
\end{equation}
\begin{equation}
\begin{gathered}
\label{Example_THO_ND1}54{\alpha}^{3}{\beta}^{6}+783{\alpha}^{2}{\beta}^{4}\mu+3222
\alpha{\beta}^{2}{\mu}^{2}+2816{\mu}^{3}+108000\alpha{\nu}^{2}\\
-4725{\alpha}^{2}{\beta}^{3}\nu-32400\alpha\beta\nu\mu \neq 0
\end{gathered}
\end{equation}
\begin{equation}
\begin{gathered}
\label{Example_THO_P2}\text{Case II}:\,
\delta=-\frac{30}{77}\,\alpha,\,
\gamma=\frac{2\alpha(990\alpha+833\beta)}{3773},\, \mu=-2\alpha
\left(\frac{\beta^2}{3}+\frac{15480720}{3195731}\alpha^2\right.\\
\left.+\frac{2916}{539}\alpha \beta\right),\,
\nu=-\alpha\left({\frac {4{\beta}^{3}}{27}}+{\frac
{88719651}{12782924}}\, {\alpha}^{2}\beta+{\frac
{26053515945}{6889996036}}\,{\alpha}^{3}+{ \frac
{4005}{1232}}\,\alpha{\beta}^{2}\right)
\end{gathered}
\end{equation}
\begin{equation}
\begin{gathered}
\label{Example_THO_ND2}\alpha\neq-\frac{5929}{5634}\,\beta,\quad
\alpha\neq\frac{5929}{103230}\,\beta,\\
{\beta}^{3}+{\frac
{104796}{4235}}\,\alpha\,{\beta}^{2}+{\frac {
5558014908}{125546575}}\,\beta\,{\alpha}^{2}+{\frac
{21448672012224}{ 1042111900445}}\,{\alpha}^{3}
 \neq 0
\end{gathered}
\end{equation}

\pr Equation \eqref{Example_THO} possesses two different asymptotic
expansions corresponding to Laurent series in a neighborhood of
poles. They are the following
\begin{equation}
\begin{gathered}
\label{Example_THO_L1}w^{(1)}(z)=-\frac{11}{\alpha\,z^2}
+\frac{11\delta-12\alpha}{19\alpha
z}+\frac{324\alpha^2-309\alpha\delta+11\delta^2+1083\alpha\beta}{8664\alpha}+\ldots
\end{gathered}
\end{equation}
\begin{equation}
\begin{gathered}
\label{Example_THO_L2}w^{(2)}(z)=\frac{6}{7z}+\left(\frac17\,\delta-
\frac{12}{49}\,\alpha-\frac13\,\beta\right)+\ldots
\end{gathered}
\end{equation}
All not written out coefficients of these series are uniquely
determined. Without loss of generality we construct Laurent series
in a neighborhood of the point $z=0$. Thus we see that equation
\eqref{Example_THO} may have an elliptic solution of order two and
an elliptic solution of order three. The second order elliptic
solution possesses one pole inside a parallelogram of periods. The
Laurent expansion in a neighborhood of the pole $z=0$ is
$w^{(1)}(z)$. While the third order elliptic solution has two
distinct poles inside a parallelogram of periods. The Laurent
expansion in a neighborhood of the pole $z=0$ is $w^{(1)}(z)$ and
the Laurent expansion in a neighborhood of the pole $z=a$ is
$w^{(2)}(z-a)$. All other elliptic solutions are obtained with a
help of transformation $z$ $\mapsto$ $z-z_0$, where $z_0$ is an
arbitrary constant. Necessary condition for the second order
elliptic solution to exist is $c_{-1}^{(1)}=0$ and we obtain the
first constrained between the parameters in expression
\eqref{Example_THO_P1}. Necessary condition for the third order
elliptic solution to exist is $c_{-1}^{(1)}+c_{-1}^{(2)}=0$.
Consequently we get the first constrained between the parameters in
expression \eqref{Example_THO_P2}. The second order elliptic
solution can be written as
\begin{equation}
\begin{gathered}
\label{Example_THO_Es1}w(z)=c_{-2}^{(1)}\wp(z;g_2,g_3)+h_0.
\end{gathered}
\end{equation}
Expanding this function in a neighborhood of the pole $z=0$, we
obtain
\begin{equation}
\begin{gathered}
\label{Example_THO_Es1_LS}w(z)=\frac{c_{-2}^{(1)}}{z^2}+h_0+\frac{g_2c_{-2}^
{(1)}}{20}z^2+\frac{g_3c_{-2}^{(1)}}{28}z^4+\ldots.
\end{gathered}
\end{equation}
Comparing coefficients of the series \eqref{Example_THO_L1} and
\eqref{Example_THO_Es1_LS}, we find the parameters $h_0$, $g_2$,
$g_3$ of elliptic solution \eqref{Example_THO_Es1}
\begin{equation}
\begin{gathered}
\label{Example_THO_Es1_P}h_0=\frac{\beta}{8},\,g_2=-\frac{\alpha(32\mu+3\beta^2\alpha)}{880},\,
g_3={\frac {{\alpha}^{2} \left(320\nu-7{\beta}^ {3}\alpha-48\beta\mu
\right)}{38720}}.
\end{gathered}
\end{equation}
and the second constraint in \eqref{Example_THO_P1}. In fact we take
seven coefficients of each series. Excluding the case of degeneracy,
i.e. $g_2^3-27g_3^2=0$, we force the condition
\eqref{Example_THO_ND1}. The third order elliptic solution is the
following
\begin{equation}
\begin{gathered}
\label{Example_THO_Es2}w(z)=c_{-2}^{(1)}\wp(z)+c_{-1}^{(1)}(\zeta(z)-
\zeta(z-a))+h_0-c_{-1}^{(1)}\zeta(a),\quad
c_{-1}^{(1)}=-c_{-1}^{(2)}
\end{gathered}
\end{equation}
where $\wp(z)\equiv$ $\wp(z;g_2,g_3)$, $\zeta(z)\equiv$
$\zeta(z;g_2,g_3)$. Expanding this function in a neighborhood of the
poles $z=0$, $z=a$ and introducing notation
$A\stackrel{def}{=}\wp(a)$, $B\stackrel{def}{=}\wp_z(a)$, we get
\begin{equation}
\begin{gathered}
\label{Example_THO_Es2_LS1}w(z)=\frac{c_{-2}^{(1)}}{z^2}+
\frac{c_{-1}^{(1)}}{z}+h_0+c_{-1}^{(1)}Az+\left\{\frac{c_{-2}^{(1)}g_2}{20}-
\frac{c_{-1}^{(1)}B}{2}\right\}z^2+\ldots.
\end{gathered}
\end{equation}
\begin{equation}
\begin{gathered}
\label{Example_THO_Es2_LS2}w(z)=
\frac{c_{-1}^{(2)}}{\xi}+h_0+c_{-2}^{(1)}A+\left\{c_{-2}^{(1)}B-c_{-1}^{(1)}A\right\}\xi
+\left\{c_{-2}^{(1)}\left(3A^2-\frac{g_2}{4}\right)\right.\\
\left.-\frac{c_{-1}^{(1)}B}{2}\right\}\xi^2+\ldots,\quad
\xi=z-a,\quad c_{-1}^{(2)}=-c_{-1}^{(1)}.
\end{gathered}
\end{equation}
Comparing coefficients of the series \eqref{Example_THO_Es2_LS1},
\eqref{Example_THO_Es2_LS2} with coefficients of $w^{(1)}(z)$ and
 $w^{(2)}(z-a)$ accordingly, we obtain an algebraic system. In this
case we take five coefficients of each series. Solving this system
together with equation $B^2=4A^3-g_2A-g_3$, we find the parameters
$h_0$, $g_2$, $g_3$, $A$, $B$
\begin{equation}
\begin{gathered}
\label{Example_THO_Es2_P}h_0={\frac
{111\alpha}{2156}}+\frac{\beta}{8},\quad g_2=\alpha^2\left({\frac
{8139}{23716}}\,\beta\alpha+{\frac {43244955}{140612164
}}\,{\alpha}^{2}+\frac{{\beta}^{2}}{48}\right)\\
g_3=-\alpha^3\left({\frac
{3163}{189728}}\,{\beta}^{2}{\alpha}+{\frac {15684093}{
562448656}}\,\beta{\alpha}^{2}+{\frac {19911334005}{1667379040712}}
\,{\alpha}^{3}+{\frac {\beta^3}{1728}}\right),\\
A=\alpha\left(\frac{\beta}{24}+\frac{69\alpha}{2156}\right),\,
B=-\alpha^2\left(\frac{15\beta}{308}+\frac{43119\alpha}{913066}\right).
\end{gathered}
\end{equation}
and the second and the third correlations in \eqref{Example_THO_P2}.
Rewriting elliptic solution \eqref{Example_THO_Es2} in terms of the
parameters $A$, $B$, yields
\begin{equation}
\begin{gathered}
\label{Example_THO_Es2_AB}w(z)=c_{-2}^{(1)}\wp(z)-\frac{c_{-1}^{(1)}(\wp_z(z)+B)}{2(\wp(z)-A)}+h_0,
\end{gathered}
\end{equation}
In order to exclude the case of degeneracy we should force the
condition \eqref{Example_THO_ND2}.

Substituting obtained elliptic solutions into equation
\eqref{Example_THO}, we see that this equation indeed possesses
solutions of the form \eqref{Example_THO_Es1},
\eqref{Example_THO_Es2_AB} provided that the parameters $\alpha$,
$\beta$, $\gamma$, $\delta$, $\mu$, $\nu$ satisfy correlations
\eqref{Example_THO_P1}, \eqref{Example_THO_P2} accordingly.

Note that if conditions \eqref{Example_THO_ND1},
\eqref{Example_THO_ND2} are not valid, then expressions
\eqref{Example_THO_Es1}, \eqref{Example_THO_Es2_AB} give
simply--periodic  or rational solutions of equation
\eqref{Example_THO} provided that correlations
\eqref{Example_THO_P1}, \eqref{Example_THO_P2} hold accordingly.

In conclusion we would like to note that solutions of equation
\eqref{Example_introduction} with $\sigma=0$ provide traveling wave
solutions $u(x,t)=b+w(z)$, $z=x-(\gamma-\delta b) t$ of the
following third order nonlinear partial differential equation
\begin{equation}
\begin{gathered}
\label{Example_THO_Partial}u_t=u_{xxx}+3\eta u u_{xx}+\eta u_x^2+a_2
u_{xx}+2\alpha u^3+\delta uu_x +a_1 u^2+a_0 u,
\end{gathered}
\end{equation}
where $a_2=\beta-3b$, $a_1=-6\alpha b$, $a_0=6\alpha b^2+\mu$ and
$b$ is a root of the equation $2\alpha b^3+\mu b-\nu=0$. Along with
this, solutions of equation \eqref{Example_introduction} give
traveling wave solutions $u(x,t)=w(z)$, $z=x-\mu t$ of the fourth
order nonlinear partial differential equation
\begin{equation}
\begin{gathered}
\label{Example_THO_Partial_1}u_t=\partial_x\left\{u_{xxx}+3\eta u
u_{xx}+\eta u_x^2+\beta u_{xx}+2\alpha u^3+\delta uu_x +\gamma
u_x+\sigma u^2\right\}.
\end{gathered}
\end{equation}
In this case the parameter $\nu$ in equation
\eqref{Example_introduction} appears as a constant of integration.

\section {Conclusion}

In this article we have studied the problem of constructing exact
elliptic solutions of autonomous nonlinear ordinary differential
equations. Our approach is based on comparing coefficients of
Laurent expansions found by asymptotic methods and coefficients of
Laurent series for the general expression of an elliptic function.
The method presented in this article generalizes several other
methods with an a priori fixed expression for an elliptic solution.
Our approach may be applied if one needs to classify elliptic
solutions of an autonomous nonlinear ordinary differential equation.

\section {Acknowledgements}

This research was partially supported by Federal Target Programm
"Research and Scientific - Pedagogical Personnel of innovation in
Russian Federation on 2009 -- 2013 (Contract P1228).


\end{document}